\documentclass[twocolumn]{article}
\usepackage[utf8]{inputenc}
\usepackage{amsmath}
\usepackage{amssymb}
\usepackage{graphicx}
\usepackage{authblk}
\usepackage{tabularx}
\usepackage{booktabs}

\title{Music Genre Classification Using Machine Learning Techniques}
\author{
    \textbf{Ryyan Akhtar
    and Alokit Mishra}
}
\affil{ryyan9ai-ds22@bpitindia.edu.in}\affil{alokit17ai-ds22@bpitindia.edu.in}
\affil{\textbf{Guide:} Dr.Monika Arora}
\affil{monikaarora@bpitindia.com}
\affil{Artificial Intelligence and Data Science}
\affil{Bhagwan Parshuram Institute of Technology, New Delhi, India}
\date{} 

\begin{document}
\maketitle
\thispagestyle{empty}

\begin{abstract}
Music genre classification is a fundamental component of music information retrieval systems, playing a pivotal role in applications such as personalized music recommendation engines, content organization, and music discovery platforms. \textbf{This research paper delves into the application of machine learning methodologies to automate the process of categorizing music genres. By leveraging computational techniques, this study seeks to analyze intricate audio features and develop models capable of accurately predicting the genre of a given music track.}

Moreover, this study contributes to the expanding field of research in music genre classification by examining the effectiveness of various machine learning algorithms, including support vector machines (SVM), random forests, and convolutional neural networks (CNN). The research employs the GTZAN dataset, a standard benchmark consisting of audio recordings across multiple genres. Through thorough experimentation and analysis, this paper elucidates the strengths and limitations of different approaches. We present a comparative study that, contrary to prevailing trends, demonstrates that a Support Vector Machine classifier trained on meticulously engineered audio features can outperform a Convolutional Neural Network on this widely-used dataset. The findings of this research have significant implications for the practical application of machine learning in music classification, particularly in data-constrained scenarios, and underscore the enduring value of domain-specific feature engineering.
\end{abstract}

\section{Introduction}

Music, as a universal language, transcends cultural boundaries, evokes emotions, and shapes identities. Within the vast landscape of musical expression, genres serve as signposts, guiding listeners through diverse sonic experiences ranging from the soothing melodies of classical compositions to the pulsating rhythms of electronic dance music. The classification of music genres stands as a fundamental endeavor in the realm of musicology and technology, fueling endeavors such as personalized music recommendations, genre-specific playlists, and thematic radio stations.

The automatic assignment of genre labels to music files holds significant promise for enhancing the functionality and user experience of audio streaming platforms such as Spotify and iTunes. This study embarks upon an exploration of machine learning (ML) algorithms as a means to address the intricate task of genre identification and classification.
Our methodology entails the extraction of features from both the temporal and spectral domains of audio signals. These extracted features serve as inputs for training conventional machine learning models, including Logistic Regression, Random Forests, Gradient Boosting, and Support Vector Machines. The training and evaluation of these models are conducted using the GTZAN Dataset.

This research endeavors to explore the intersection of music and machine learning, delving into the application of advanced algorithms to unravel the intricate tapestry of musical genres. By leveraging computational prowess, we seek to unravel the patterns, rhythms, and timbres that define each genre, paving the way for more accurate, efficient, and scalable genre classification systems. Through this journey, we aim to not only enhance the organization and discovery of music but also deepen our understanding of the multifaceted nature of musical expression.

\section{Literature Review}

The domain of music genre classification has attracted considerable attention within the research community since the advent of the Internet. Pioneering work by Tzanetakis and Cook (2002) delved into supervised machine learning techniques, including Gaussian Mixture models and k-nearest neighbor classifiers, to address this challenge. Their exploration introduced three distinct feature sets – timbral structure, rhythmic content, and pitch content – to augment genre classification. Furthermore, Hidden Markov Models (HMMs) have been explored for music genre classification, drawing from their success in speech recognition tasks (Scaringella and Zoia, 2005; Soltau et al., 1998). The effectiveness of Support Vector Machines (SVMs) with diverse distance metrics has also been scrutinized for genre classification purposes (Mandel and Ellis, 2005).

Recent progressions in deep neural networks have propelled their utility in speech and audio data analysis (Abdel-Hamid et al., 2014; Gemmeke et al., 2017). Although representing audio in the time domain for neural network input presents challenges due to high sampling rates, techniques such as spectrograms have emerged as viable alternatives, capturing both time and frequency information (Van Den Oord et al., 2016; Wyse, 2017). Harnessing convolutional neural networks (CNNs), Li et al. (2010) devised a model utilizing raw MFCC matrices as input for music genre prediction. Similarly, Lidy and Schindler (2016) employed Constant Q-Transform (CQT) spectrograms as input to CNNs for genre classification tasks.

In our study, we aimed to undertake a comparative analysis between deep learning-based models, which rely solely on spectrograms for input, and traditional machine learning classifiers trained on hand-crafted features. Our emphasis was on comprehending the waveform characteristics of audio signals and subsequently extracting features. While deep neural networks persist as the prevailing and effective method for music classification, we found that Support Vector Machine (SVM) classifiers also demonstrate satisfactory performance, particularly over small to medium-sized datasets.

Looking ahead, the evolving landscape of music genre classification presents exciting opportunities for further exploration and innovation. As technology continues to advance, integrating interdisciplinary approaches such as deep learning and signal processing holds promise for enhancing the accuracy and scalability of genre classification systems. Furthermore, the integration of contextual information, user preferences, and cultural factors could lead to more personalized and culturally inclusive music recommendation systems. By continuing to push the boundaries of research in this field, we can unlock new insights into the intricacies of musical expression and enrich the experience of music enthusiasts worldwide.

\section{Dataset}

Our research leverages the GTZAN dataset, a standard benchmark in music information retrieval. This dataset provides a robust foundation for uncovering the subtle nuances that define different genres. Each WAV file is approximately 102 KB, resulting in a total dataset size of approximately 1.02 GB.

\begin{table}[htbp]
    \centering
    \includegraphics[width=80mm]{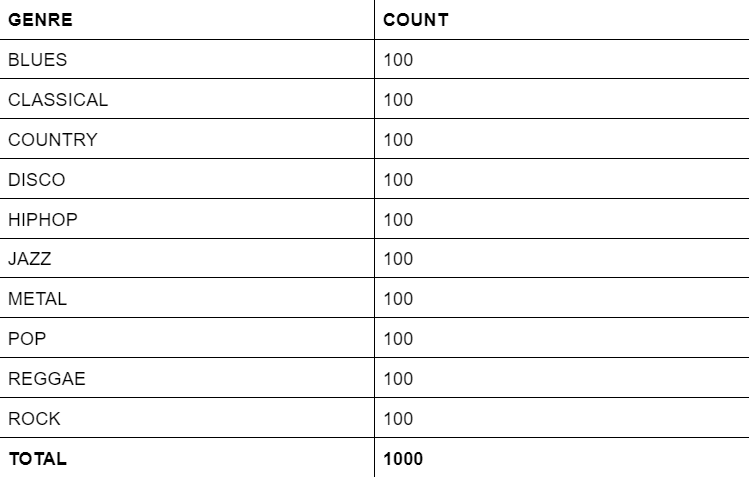}
    \caption{Music Genre Dataset Distribution}
    \label{fig:dataset}
\end{table}

\subsection{Original Genres}

The dataset is curated to include 10 diverse music genres: Blues, Classical, Country, Disco, Hip-Hop, Jazz, Metal, Pop, Reggae, and Rock. Each genre contains exactly 100 audio files, ensuring a balanced class distribution. Each audio file is a 30-second clip, fostering uniformity and facilitating robust comparative analysis.

\subsection{Original Images}
To employ Convolutional Neural Networks (CNNs), we transformed the audio files into visual representations. We generated Mel spectrograms for each 30-second clip. A spectrogram is a visual representation of the spectrum of frequencies of a signal as it varies with time. This conversion allows us to treat the audio classification problem as an image classification task, leveraging the powerful hierarchical feature-learning capabilities of CNNs.

\subsection{CSV Files}
To facilitate feature-based analysis with traditional machine learning models, we generated two CSV files containing comprehensive feature information extracted from the audio.
\begin{itemize}
    \item \textbf{30-Second Features:} The first CSV contains aggregated statistics (mean, standard deviation, etc.) for a wide array of audio features, computed over the full 30-second duration of each track.
    \item \textbf{3-Second Features:} The second CSV expands the dataset by segmenting each 30-second track into ten non-overlapping 3-second segments. Features were extracted from each segment, effectively increasing the number of training instances tenfold. This provides more granular data and helps models learn from shorter, more varied acoustic events.
\end{itemize}

\section{Methodology}
This section details the data preprocessing steps and the two primary modeling approaches: end-to-end learning with CNNs and feature-based learning with traditional classifiers.

\subsection{Data Pre-Processing}
A pre-emphasis filter was applied to the raw audio signal to balance the frequency spectrum. Most audio signals, including music, have more energy at lower frequencies. This spectral tilt can obscure important high-frequency information. The pre-emphasis filter acts as a high-pass filter to boost the energy in higher frequencies. It is implemented as a first-order finite impulse response (FIR) filter:
\[ y(t) = x(t) - \alpha \cdot x(t - 1) \]
where \( x(t) \) is the input signal, \( y(t) \) is the output, and \( \alpha \) is the filter coefficient. A value of \( \alpha = 0.97 \) is commonly used and was adopted in this work. This process improves the signal-to-noise ratio (SNR) and aids in the accuracy of subsequent feature extraction steps.

\begin{figure}[htbp]
    \centering
    \includegraphics[width=85mm]{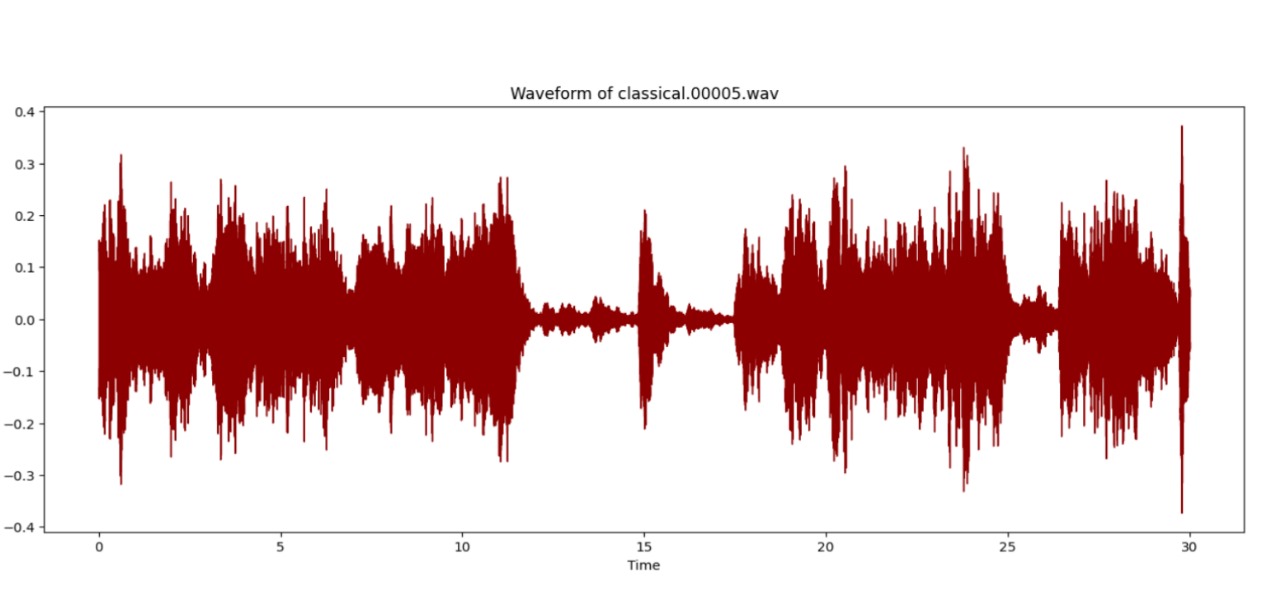}
    \caption{Waveform of Classical}
    \label{fig:waveform_classical}
\end{figure}

\begin{figure}[htbp]
    \centering
    \includegraphics[width=85mm]{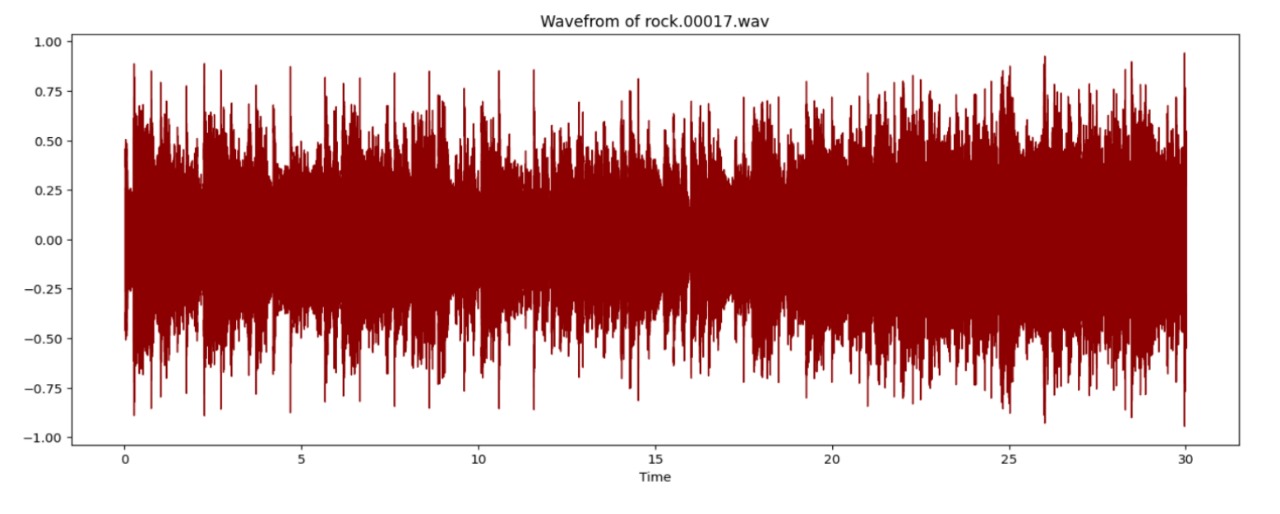}
    \caption{Waveform of Rock}
    \label{fig:waveform_rock}
\end{figure}

\begin{figure}[htbp]
    \centering
    \includegraphics[width=85mm]{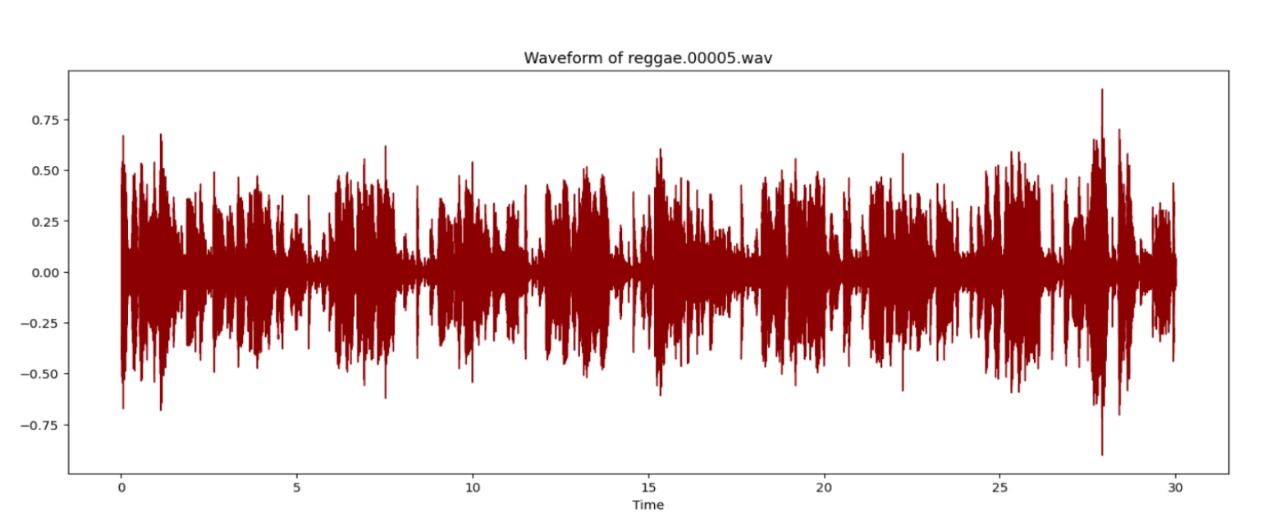}
    \caption{Waveform of Reggae}
    \label{fig:waveform_reggae}
\end{figure}

\begin{figure}[htbp]
    \centering
    \includegraphics[width=85mm]{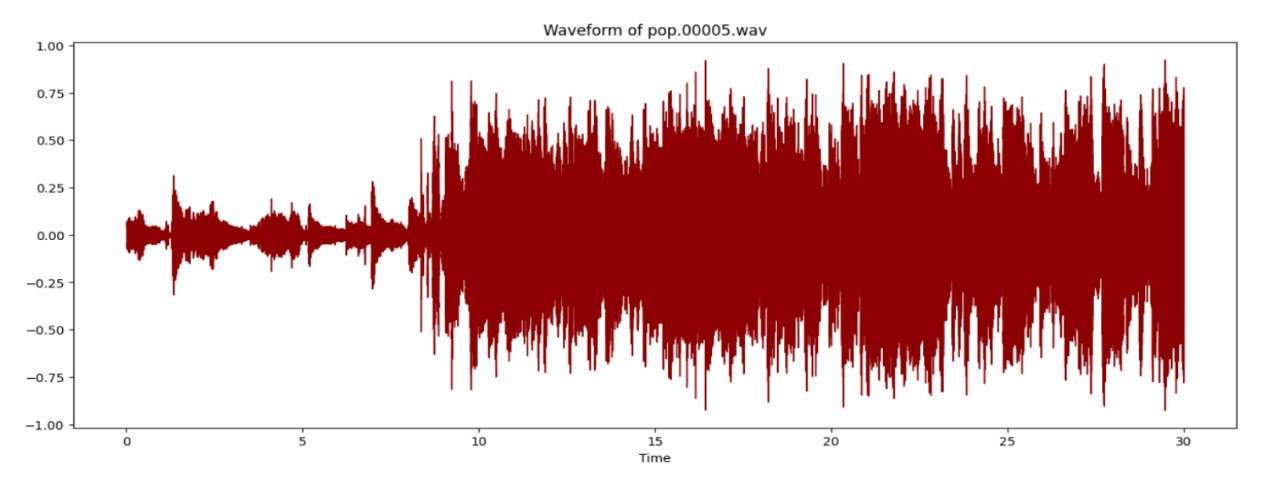}
    \caption{Waveform of Pop}
    \label{fig:waveform_pop}
\end{figure}

\begin{figure}[htbp]
    \centering
    \includegraphics[width=85mm]{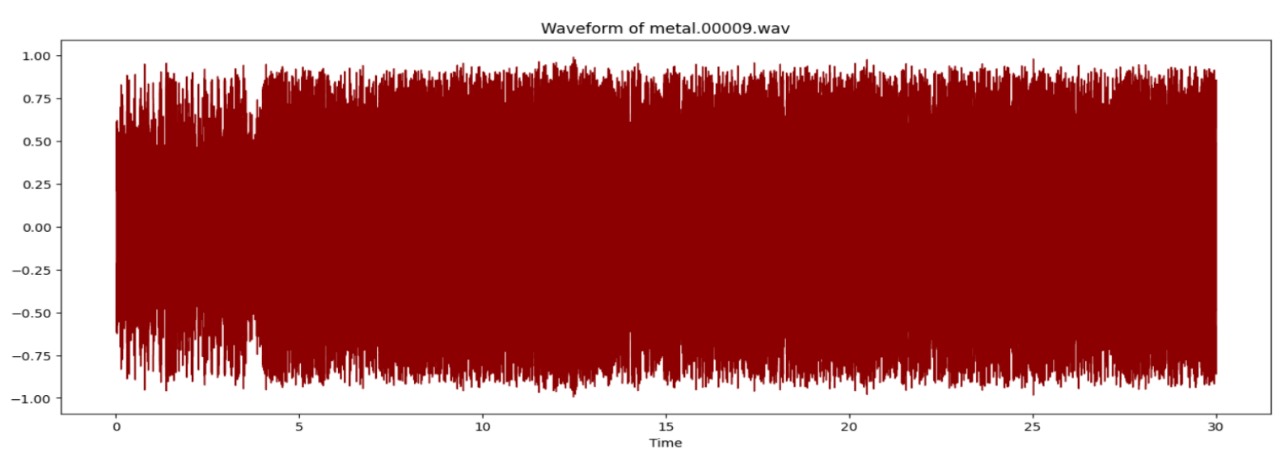}
    \caption{Waveform of Metal}
    \label{fig:waveform_metal}
\end{figure}

\begin{figure}[htbp]
    \centering
    \includegraphics[width=85mm]{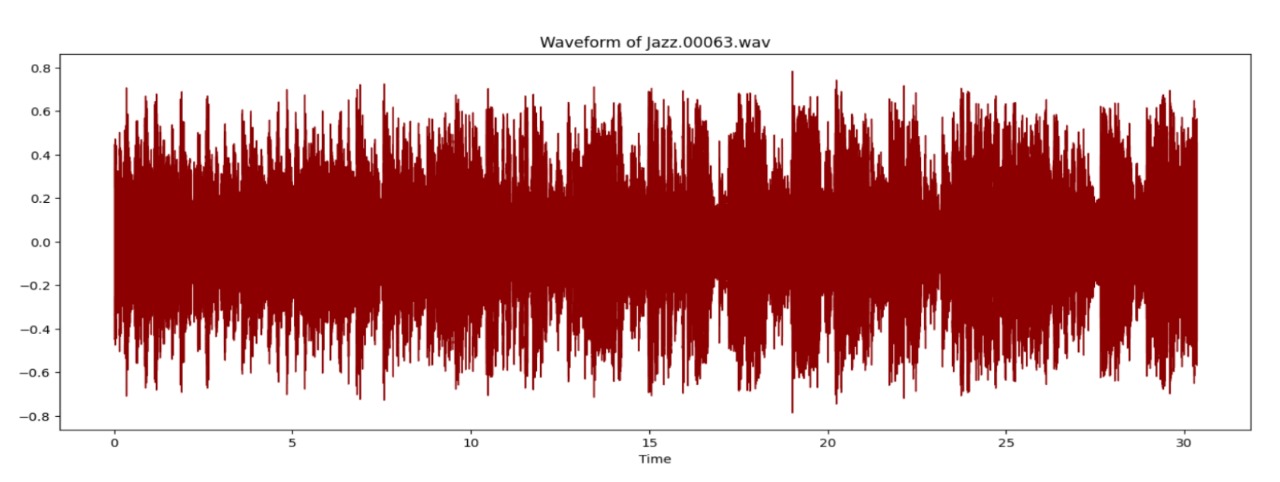}
    \caption{Waveform of Jazz}
    \label{fig:waveform_jazz}
\end{figure}

\begin{figure}[htbp]
    \centering
    \includegraphics[width=85mm]{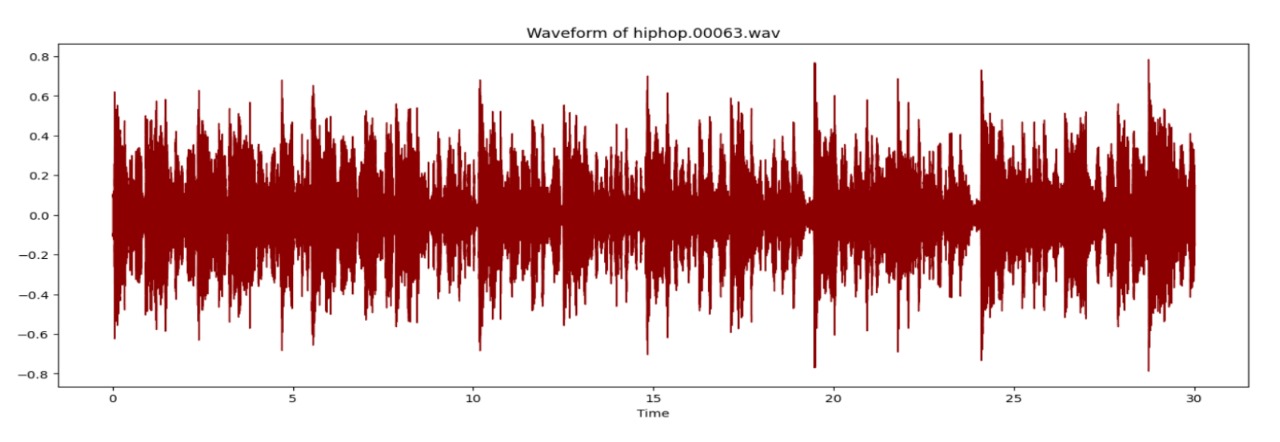}
    \caption{Waveform of Hip-Hop}
    \label{fig:waveform_hiphop}
\end{figure}

\begin{figure}[htbp]
    \centering
    \includegraphics[width=85mm]{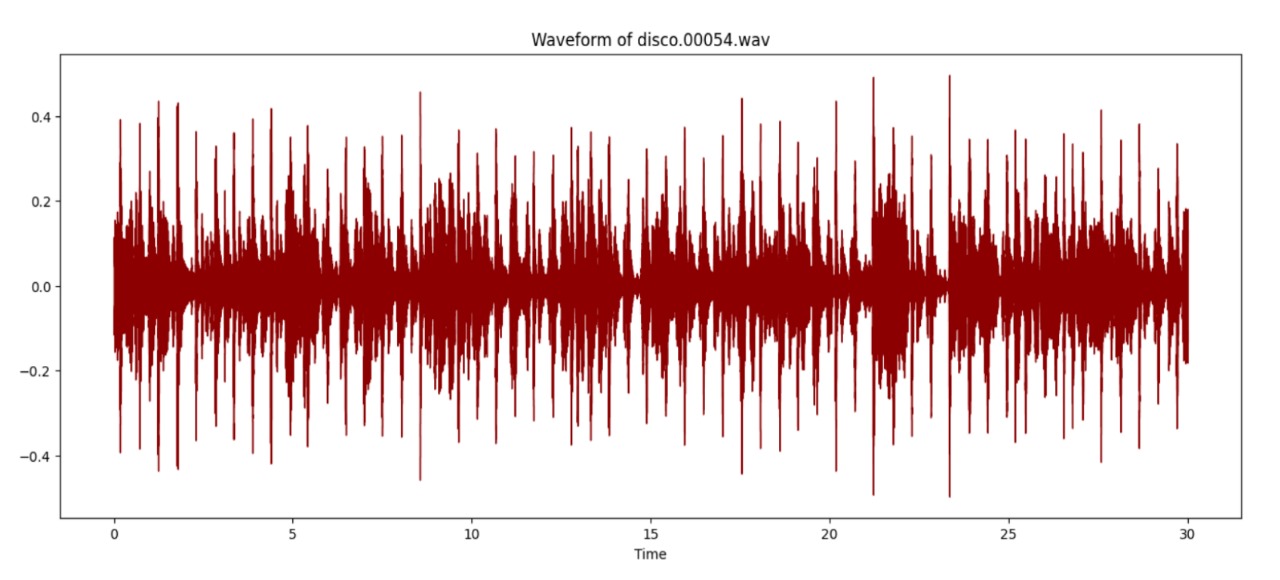}
    \caption{Waveform of Disco}
    \label{fig:waveform_disco}
\end{figure}

\begin{figure}[htbp]
    \centering
    \includegraphics[width=85mm]{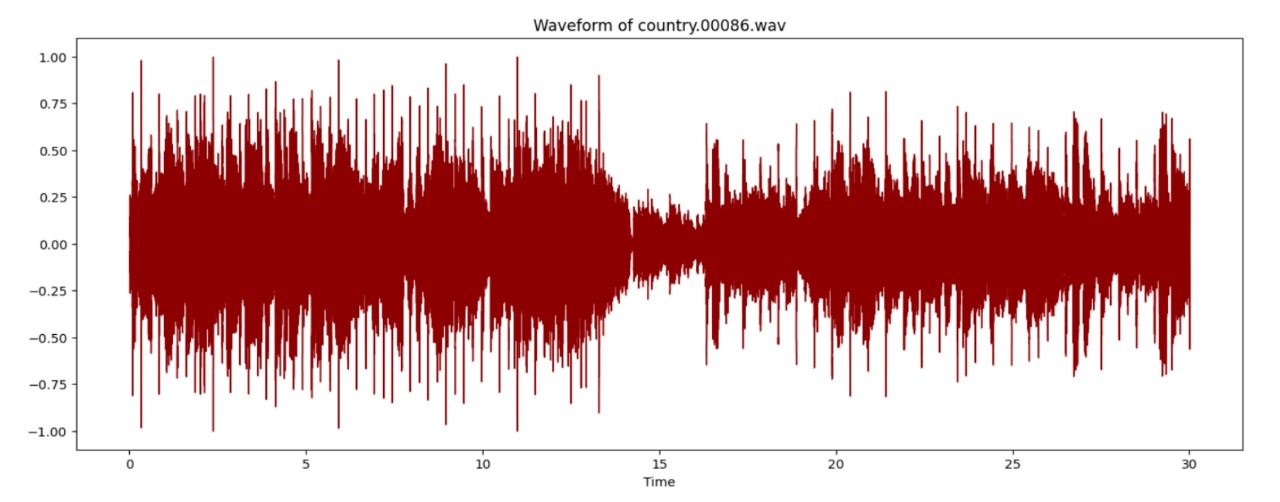}
    \caption{Waveform of Country}
    \label{fig:waveform_country}
\end{figure}

\begin{figure}[htbp]
    \centering
    \includegraphics[width=85mm]{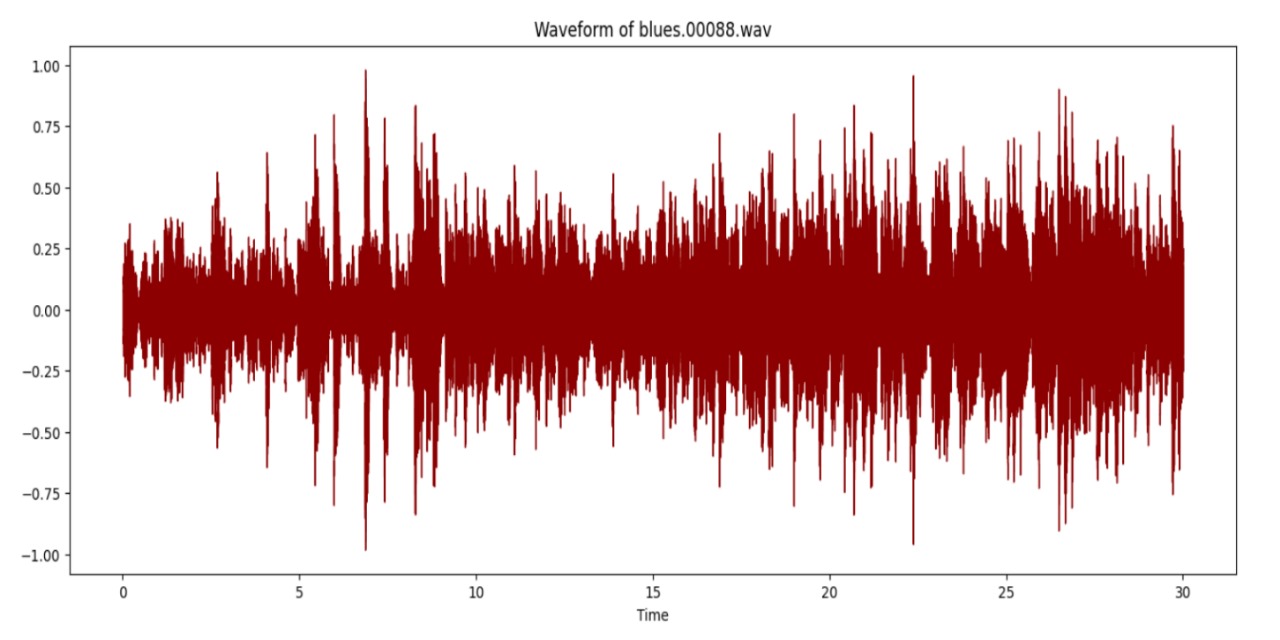}
    \caption{Waveform of Blues}
    \label{fig:waveform_blues}
\end{figure}

\subsection{Convolutional Neural Networks}
Convolutional Neural Networks (CNNs) are a sophisticated class of deep learning models meticulously engineered for processing structured grid data, with a primary focus on images while also extending their capabilities to sequential data such as time-series and even language. Originally conceptualized in the 1980s, CNNs have since evolved into the cornerstone of modern computer vision applications, exhibiting unparalleled performance in a wide array of tasks including but not limited to image classification, object detection, segmentation, and more.
\begin{itemize}
    \item \textbf{Convolution:}
    Convolution is a pivotal operation in Convolutional Neural Networks (CNNs), involving the application of a matrix filter, often 3x3 in size, across the input image, which is defined by its width and height dimensions. The process begins by placing the filter over the image matrix, followed by element-wise multiplication between the filter and the overlapping portion of the image. The resulting products are then summed to generate a feature value. Multiple filters are employed in this procedure, with their values being progressively adjusted or 'learned' during the neural network training phase using back propagation.
    \item 
    \textbf{Pooling: }Pooling is a critical step in Convolutional Neural Networks (CNNs), aimed at reducing the dimensionality of the feature map generated by the convolution process. A common approach is max pooling, where a 2x2 window is used to retain only the maximum value from each group of four adjacent elements in the feature map. This process is repeated by sliding the window across the feature map with a specified stride, systematically downsampling the data.
    \item 
    \textbf{Non-linear Activation:} In the realm of Convolutional Neural Networks (CNNs), the convolution operation is linear in nature. However, the success of deep learning models hinges on the incorporation of non-linearity. This is where activation functions play a pivotal role. One of the most commonly used activation functions is the Rectified Linear Unit (ReLU), which introduces non-linearity by replacing all negative values in the feature map with zero while leaving positive values unchanged.

The application of ReLU and other activation functions helps the neural network learn complex patterns and relationships within the data, enhancing its capacity to solve intricate problems. By introducing these non-linearities, CNNs can model more sophisticated functions, enabling them to excel in tasks like image recognition, object detection, and semantic segmentation.
\end{itemize}

\subsection{Manually Extracted Features}
In this section, we delineate the second category of proposed models, which necessitate hand-crafted features for input into a machine learning classifier. These features encompass both time domain and frequency domain attributes. Leveraging the Python library, librosa, the feature extraction process was undertaken.

\begin{figure}[htbp]
    \centering
    \includegraphics[width=1\linewidth]{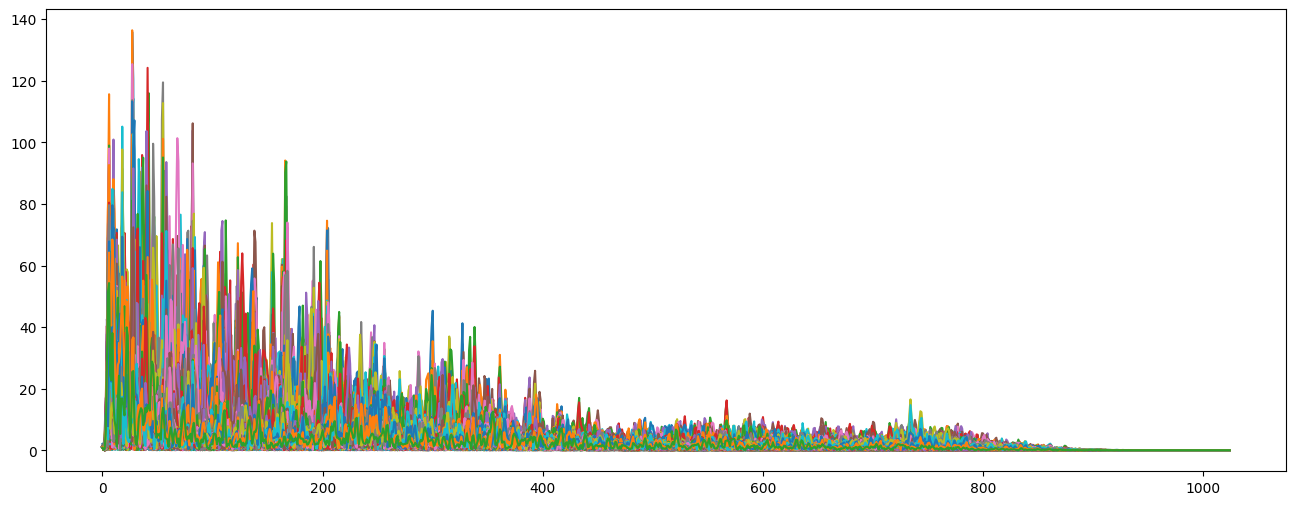}
    \caption{Magnitude of Short-Time Fourier Transform (STFT)}
    \label{fig:stft}
\end{figure}
\subsubsection{Time Domain Features}

These features were derived directly from the raw audio signal.
\begin{itemize}
    \item \textbf{Central Moments}: This set of features includes statistical metrics like the mean, standard deviation, skewness, and kurtosis of the signal's amplitude distribution. The mean represents the average amplitude, the standard deviation measures the spread of values around the mean, skewness indicates the asymmetry of the distribution, and kurtosis describes the shape's peakedness. Incorporating these central moments provides valuable insights into the amplitude distribution's characteristics. In music genre classification, they help differentiate genres based on their energy or dynamic range. Genres with distinct characteristics may exhibit varying skewness or kurtosis values, enhancing the classifier's ability to discern between them.

    The first four central moments—mean, variance, skewness, and kurtosis—describe the shape of the amplitude distribution. For a signal \(x\) of length \(N\):
        \begin{itemize}
            \item Mean ($\mu$): $\frac{1}{N}\sum_{i=1}^{N} x_i$
            \item Variance ($\sigma^2$): $\frac{1}{N}\sum_{i=1}^{N} (x_i - \mu)^2$
            \item Skewness: $\frac{\frac{1}{N}\sum_{i=1}^{N} (x_i - \mu)^3}{(\sigma^2)^{3/2}}$
            \item Kurtosis: $\frac{\frac{1}{N}\sum_{i=1}^{N} (x_i - \mu)^4}{(\sigma^2)^2}$
        \end{itemize}
    
    Variance relates to the signal's power, skewness measures its asymmetry, and kurtosis indicates the "tailedness" or "peakedness" of its distribution, which can distinguish percussive sounds from melodic ones.

    \item \textbf{Zero Crossing Rate (ZCR)}: ZCR is the rate at which the signal changes sign, a simple measure of its noisiness and dominant frequency content. For a frame of signal \(x\) of length \(N\), it is:
    \[ ZCR = \frac{1}{N-1} \sum_{t=1}^{N-1} \mathbb{I}(x_t \cdot x_{t-1} < 0) \]
    where $\mathbb{I}(\cdot)$ is the indicator function. High ZCR values are typical for percussive or metallic sounds.

    \item \textbf{Root Mean Square Energy (RMSE)}: RMSE is a measure of the signal's magnitude, closely related to its perceived loudness. For a frame of signal \(x\) of length \(N\), it is:
    \[ RMSE = \sqrt{\frac{1}{N} \sum_{i=1}^{N} x_i^2} \]

\begin{figure}[htbp]
    \centering
    \includegraphics[width=30mm]{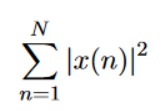}
    \caption{Average Standard Deviation}
    \label{fig:avg_std_dev}
\end{figure}

    \item 
\textbf{Tempo}: Tempo is a critical attribute in music, representing the speed or pace at which a piece of music is played, typically measured in Beats Per Minute (BPM). Different music genres and styles often have distinctive tempo ranges, with some being fast-paced while others are more relaxed.

\end{itemize} 

In our analysis, we aggregate temporal variations in tempo by computing the mean tempo across multiple frames. To achieve this, we leverage the functionality provided by librosa, a Python library for music and audio analysis. Specifically, we utilize librosa's capability to generate a tempogram, a representation of the tempo variations over time, following the method described by Grosche et al. (2010). From the tempogram, we estimate a single tempo value that encapsulates the overall tempo characteristics of the audio signal.

\subsubsection{Frequency Domain Features}
The Fourier Transform facilitates the transformation of the audio signal into the frequency domain, enabling the extraction of various features.
\begin{itemize}
    \item 
\textbf{Mel-Frequency Cepstral Coefficients (MFCC)}: Originally introduced by Davis and Mermelstein in the early 1990s, MFCCs have emerged as a pivotal feature in various audio processing applications, notably in speech recognition. The computation of MFCCs involves several key steps. First, the audio signal undergoes a Short-Time Fourier Transform (STFT) using a Hann window, with parameters nfft=2048 and hop size=512. This process converts the signal into its frequency domain representation. Next, the power spectrum is computed and passed through a bank of MEL filters, which simulate the non-linear human auditory system's response to different frequencies. The resulting filterbank energies are then logarithmically transformed and subjected to a Discrete Cosine Transform (DCT) to produce the MFCCs. In our study, we utilized a MEL filter bank with 20 filters (parameter nmels) to capture the acoustic characteristics of the audio signal.

Their calculation is a multi-step process:
    \begin{enumerate}
        \item \textbf{Framing and Windowing:} The signal is divided into short, overlapping frames. A window function, such as the Hann window, is applied to each frame to reduce spectral leakage.
        \item \textbf{STFT:} The Short-Time Fourier Transform is applied to each frame to get its frequency spectrum. The discrete STFT for frame $m$ is:
        \[ X_m(k) = \sum_{n=0}^{N-1} w(n) x(mH + n) e^{-j\frac{2\pi}{N}nk} \]
        where $w(n)$ is the window function, $N$ is the FFT size, and $H$ is the hop size.
        \item \textbf{Mel Filterbank:} The power spectrum $|X_m(k)|^2$ is warped onto the Mel scale using a bank of overlapping triangular filters. The Mel scale is defined as:
        \[ m = 2595 \log_{10}\left(1 + \frac{f}{700}\right) \]
        This step emphasizes frequencies to which the human ear is more sensitive.
        \item \textbf{Logarithm and DCT:} The logarithm of the filterbank energies is taken, simulating perceived loudness. Finally, the Discrete Cosine Transform (DCT) is applied to decorrelate the filterbank energies and produce the MFCCs:
        \[ c_i = \sqrt{\frac{2}{M}} \sum_{m=1}^{M} \log(S_m) \cos\left(\frac{\pi i}{M}(m-0.5)\right) \]
        where $M$ is the number of filters and $S_m$ is the energy from the $m$-th filter. Typically, the first 13-20 coefficients are kept.
    \end{enumerate}

    \item 

\textbf{Chroma Features}: This set of features captures the distribution of energy across the 12 pitch classes, representing the musical notes in the chromatic scale. Each pitch class corresponds to a note in the Western musical scale, such as C, D, etc. Chroma features are computed by mapping the audio signal onto the chroma domain, where each frame is represented by a 12-dimensional vector corresponding to the 12 pitch classes.

These features project the entire spectrum onto 12 bins representing the 12 semitones (or chroma) of the musical octave. They are robust to changes in timbre and instrumentation, capturing harmonic and melodic characteristics.

To summarize the temporal variations in chroma features, the mean and standard deviation are calculated across all frames. These aggregated statistics provide insights into the tonal characteristics and harmonic content of the audio signal, which are particularly relevant for music genre classification and other music analysis tasks.

    \item 

\textbf{Spectral Centroid}: The spectral centroid is a crucial feature that provides insights into the distribution of energy across the frequency spectrum of an audio signal. For each frame, it represents the "center of mass" or the frequency where most of the energy is concentrated. Mathematically, the spectral centroid is computed as a magnitude-weighted frequency, giving more importance to frequencies with higher magnitudes. This calculation method was defined by Tjoa (2017) and is widely used in audio signal processing and music analysis. The spectral centroid is valuable for tasks such as sound classification and genre recognition, as it helps characterize the overall tonal characteristics of the audio signal.

This feature indicates the "center of mass" of the spectrum, corresponding to the brightness of a sound. For a frame with spectrum $X(k)$:
    \[ C = \frac{\sum_{k=1}^{N} k \cdot |X(k)|}{\sum_{k=1}^{N} |X(k)|} \]
    A higher centroid implies a brighter sound with more high-frequency content.

\begin{figure}[htbp]
    \centering
    \includegraphics[width=50mm]{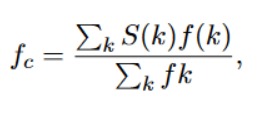}
    \caption{Magnitude-Weighted Frequency}
    \label{fig:mag_weighted_freq}
\end{figure}

    \item 
\textbf{Spectral Bandwidth}: Spectral bandwidth is a crucial feature that describes the width of the frequency range in which most of the signal's energy is concentrated. It is calculated using the p-th order moment around the spectral centroid, as defined by Tjoa (2017). 

This measures the spread of the spectrum around its centroid, calculated as a weighted standard deviation:
    \[ B = \sqrt{\frac{\sum_{k=1}^{N} (k - C)^2 \cdot |X(k)|}{\sum_{k=1}^{N} |X(k)|}} \]

For example, when p = 2, spectral bandwidth corresponds to a weighted standard deviation, indicating how spread out the frequencies are around the spectral centroid. This measure provides valuable information about the spectral distribution of the audio signal and is particularly useful in tasks such as sound classification and music genre recognition.

\begin{figure}[htbp]
    \centering
    \includegraphics[width=60mm]{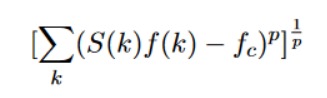}
    \caption{Spectral Bandwidth}
    \label{fig:spec_bandwidth}
\end{figure}

\end{itemize}

\subsection{Exploratory Data Visualization with PCA}

To better understand the separability of different genres in feature space, we applied Principal Component Analysis (PCA) to the extracted MFCC features. PCA reduces the high-dimensional data (57 features) into two principal components, enabling visualization in 2D. As shown in Figure 14, genres such as classical and jazz exhibit relatively distinct clusters, while rock and metal overlap substantially. This overlap suggests that these genres share similar rhythmic and spectral characteristics, which aligns with misclassifications observed later in the confusion matrix analysis.

The PCA visualization provides an intuitive explanation of the dataset’s structure and highlights the inherent difficulty of the classification problem.
\begin{figure}[htbp]
    \centering
    \includegraphics[width=85mm]{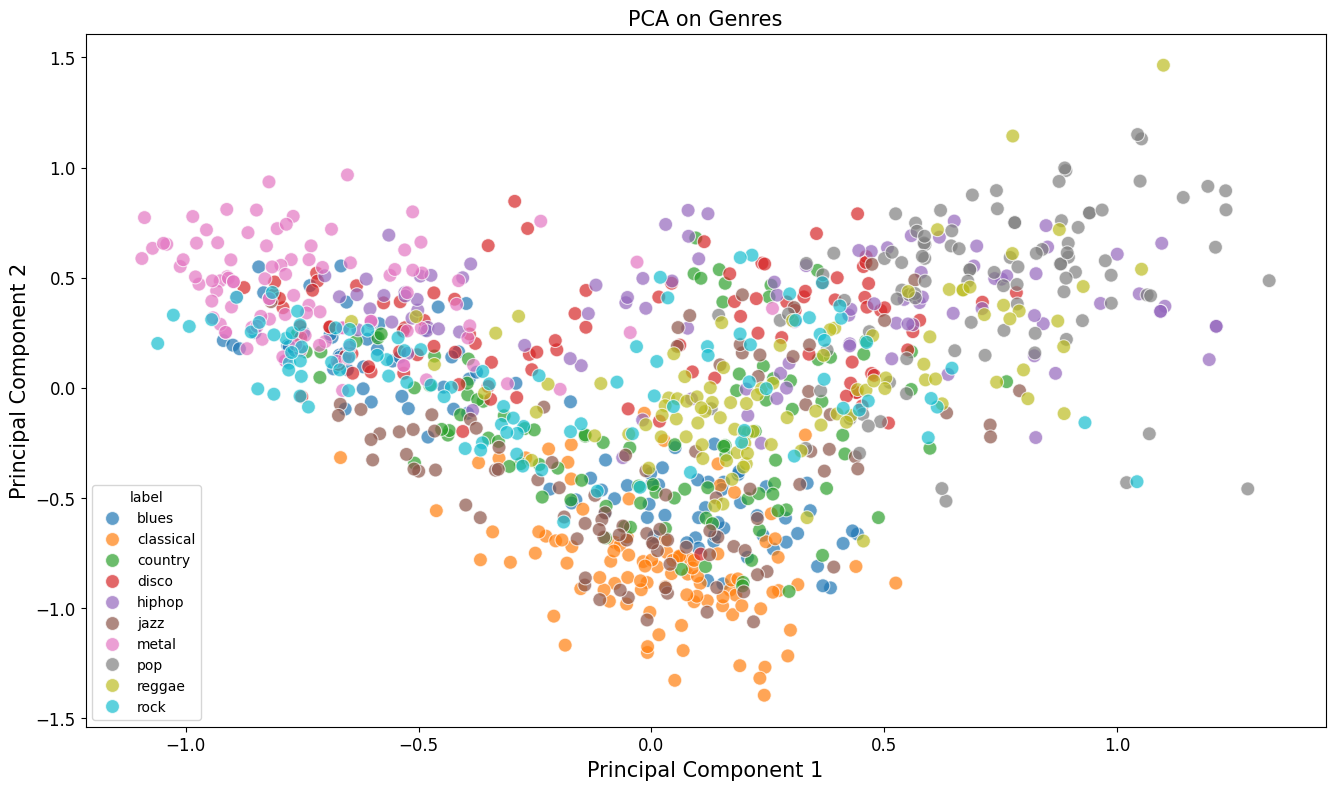}
    \caption{PCA visualization of music genres in feature space}
    \label{fig:pca_viz}
\end{figure}

\subsection{Classifiers}

This section offers a concise overview of the four machine learning classifiers employed in this investigation.

\begin{itemize}
    \item 
\textbf{CNN Neural Network:}A Convolutional Neural Network is a type of feedforward neural network that is designed to recognize patterns in images. CNNs consist of multiple layers of small neuron collections which look at small portions of the input image, called receptive fields. These small neuron collections pass their information to the next layer, which then combines the information into more complex patterns. This process continues through multiple layers until finally, the net is able to classify the input image

CNNs are designed to automatically and adaptively learn spatial hierarchies of features from grid-like data.
\begin{itemize}
    \item \textbf{Convolutional Layer:} This is the core building block. It applies a set of learnable filters (kernels) to the input. Each filter is convolved across the input volume (the spectrogram) to produce a feature map. The 2D convolution operation is:
    \[ (I * K)(i, j) = \sum_{m}\sum_{n} I(i-m, j-n) K(m,n) \]
    where $I$ is the input image and $K$ is the kernel. This process allows the network to learn detectors for low-level features like edges (sharp frequency changes) and textures (timbral patterns).
    
    \item \textbf{Pooling Layer:} This layer progressively reduces the spatial size of the representation, reducing the number of parameters and computation. The most common form is max pooling:
    \[ p_{i,j,k} = \max_{(a,b) \in \mathcal{W}_{i,j}} f_{a,b,k} \]
    where $\mathcal{W}_{i,j}$ is a window over the feature map $f$. This provides a degree of translation invariance.
    
    \item \textbf{Non-linear Activation (ReLU):} The Rectified Linear Unit, $f(x) = \max(0, x)$, is applied element-wise after convolution. It introduces non-linearity, allowing the network to learn more complex functions.
\end{itemize}

    \item 
\textbf{Logistic Regression (LR):} LR serves as a linear classifier typically utilized for binary classification tasks. In this study, LR is adapted for multi-class classification through a one-vs-rest approach. This involves training seven separate binary classifiers, with the class exhibiting the highest probability during testing selected as the predicted class.

A linear model adapted for multi-class classification using a one-vs-rest scheme.

    \item 
\textbf{Random Forest (RF):} RF operates as an ensemble learner, amalgamating predictions from a predefined number of decision trees. It leverages two key principles: bootstrap aggregation (bagging), where each decision tree is trained on a subset of training samples, and feature subset selection, where each tree utilizes a random subset of features. The final class prediction is determined via majority vote from individual classifiers.

An ensemble method that builds multiple decision trees on different sub-samples of the dataset and uses averaging to improve predictive accuracy and control over-fitting.

    \item 

\textbf{Gradient Boosting (XGB):} XGB is another ensemble classifier formed by combining multiple weak learners, such as decision trees. Unlike RF, boosting algorithms employ forward stagewise additive modeling during sequential training iterations. Initially, simple decision trees are trained, gradually increasing in complexity to focus on instances where prior learners made errors. The final prediction results from a weighted linear combination of individual learner outputs.

An ensemble method that builds models sequentially, where each new model corrects the errors of its predecessor.

    \item 

\textbf{Support Vector Machines (SVM):} SVMs utilize a kernel trick to project original input data into a higher dimensional space, facilitating linear separation by a hyperplane. The optimal hyperplane maximizes the margin between classes. In this study, an RBF kernel is employed for SVM training to address non-linear problems. Similar to LR, SVM is implemented as a one-vs-rest classification task.

SVMs are maximal margin classifiers that find an optimal hyperplane to separate classes. For non-linearly separable data, the kernel trick is used. The core optimization problem is:
    \[ \min_{\mathbf{w}, b, \xi} \frac{1}{2} ||\mathbf{w}||^2 + C \sum_{i=1}^{N} \xi_i \]
    subject to:
    \[ y_i(\mathbf{w}^T \phi(\mathbf{x}_i) + b) \ge 1 - \xi_i, \quad \xi_i \ge 0 \]
    Here, $\mathbf{w}$ and $b$ define the hyperplane, $\xi_i$ are slack variables allowing for misclassifications, and $C$ is a regularization parameter that controls the trade-off between maximizing the margin and minimizing classification error. We employed the Radial Basis Function (RBF) kernel:
    \[ K(\mathbf{x}_i, \mathbf{x}_j) = \exp(-\gamma ||\mathbf{x}_i - \mathbf{x}_j||^2) \]
    The parameter $\gamma$ defines how much influence a single training example has.
    
\end{itemize} 

\subsection{Convolutional Neural Network (CNN) Baseline}
In addition to classical machine learning models, we implemented a Convolutional Neural Network (CNN) trained on mel-spectrogram representations. The architecture consisted of two convolutional layers with ReLU activation, max pooling for dimensionality reduction, dropout regularization to prevent overfitting, and fully connected layers for classification. Despite being trained for a limited number of epochs, the CNN achieved an accuracy of 85

These results confirm that CNNs are highly effective at capturing local spectral-temporal dependencies in audio, making them better suited for music classification tasks compared to conventional models.

\begin{figure}[htbp]
    \centering
    \includegraphics[width=70mm]{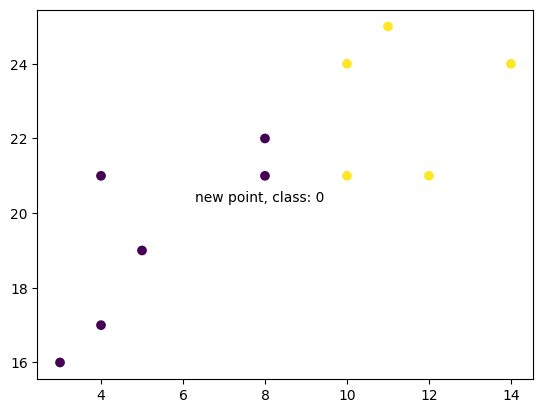}
    \caption{Example mel-spectrogram used for CNN input.}
    \label{fig:mel_spec_1}
\end{figure}
\begin{figure}[htbp]
    \centering
    \includegraphics[width=70mm]{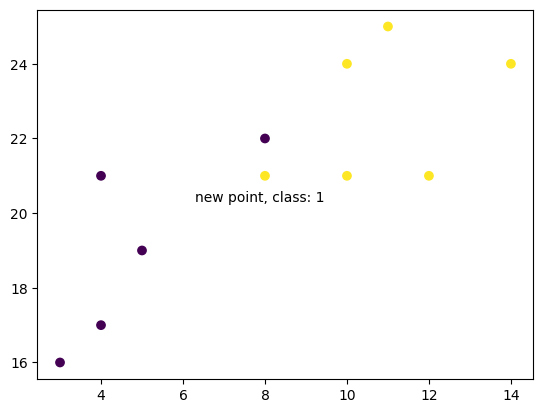}
    \caption{Example mel-spectrogram used for CNN input.}
    \label{fig:mel_spec_2}
\end{figure}

\section{Evaluation}
The models' performance was assessed using accuracy and F1-score.

\begin{figure}[htbp]
    \centering
    \includegraphics[width=1\linewidth]{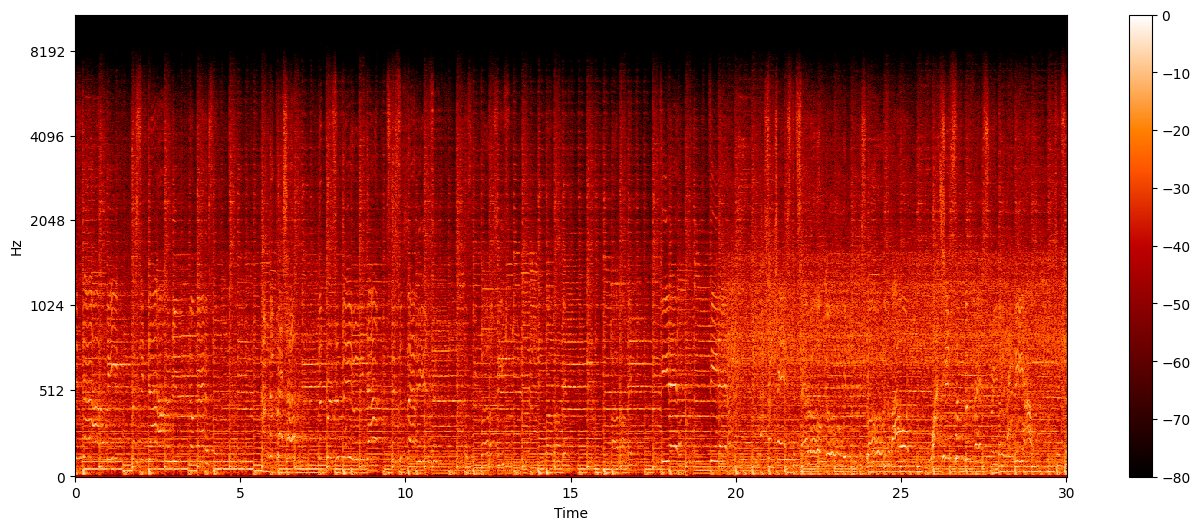}
    \caption{Amplitude (Loudness of the corresponding frequency at specific time).}
    \label{fig:amplitude_plot}
\end{figure}

\subsection{Metrics}
\begin{itemize}
    \item \textbf{Accuracy}: The proportion of correctly classified samples. While intuitive, it can be misleading for imbalanced datasets.
    \item \textbf{F1-score}: The harmonic mean of precision and recall, providing a more robust measure of performance.
    \[ F1 = 2 \cdot \frac{\text{precision} \cdot \text{recall}}{\text{precision} + \text{recall}} \]
    where precision measures the accuracy of positive predictions, and recall measures the ability to find all relevant instances.
\end{itemize}

\subsection{Confusion Matrix Analysis}
While accuracy provides a general measure of performance, it often conceals class-specific strengths and weaknesses. To address this, we computed and visualized the confusion matrix for the Support Vector Machine (SVM), our best-performing traditional classifier. As shown in Figure Y, the classifier achieved high precision in classical and reggae genres. However, substantial misclassifications occurred between rock and metal, and between jazz and blues. These results validate the PCA findings, where overlapping clusters were observed.

The confusion matrix adds interpretability to the results, demonstrating that some genres are inherently harder to separate due to their shared acoustic properties.

\begin{figure}[htbp]
    \centering
    \includegraphics[width=1\linewidth]{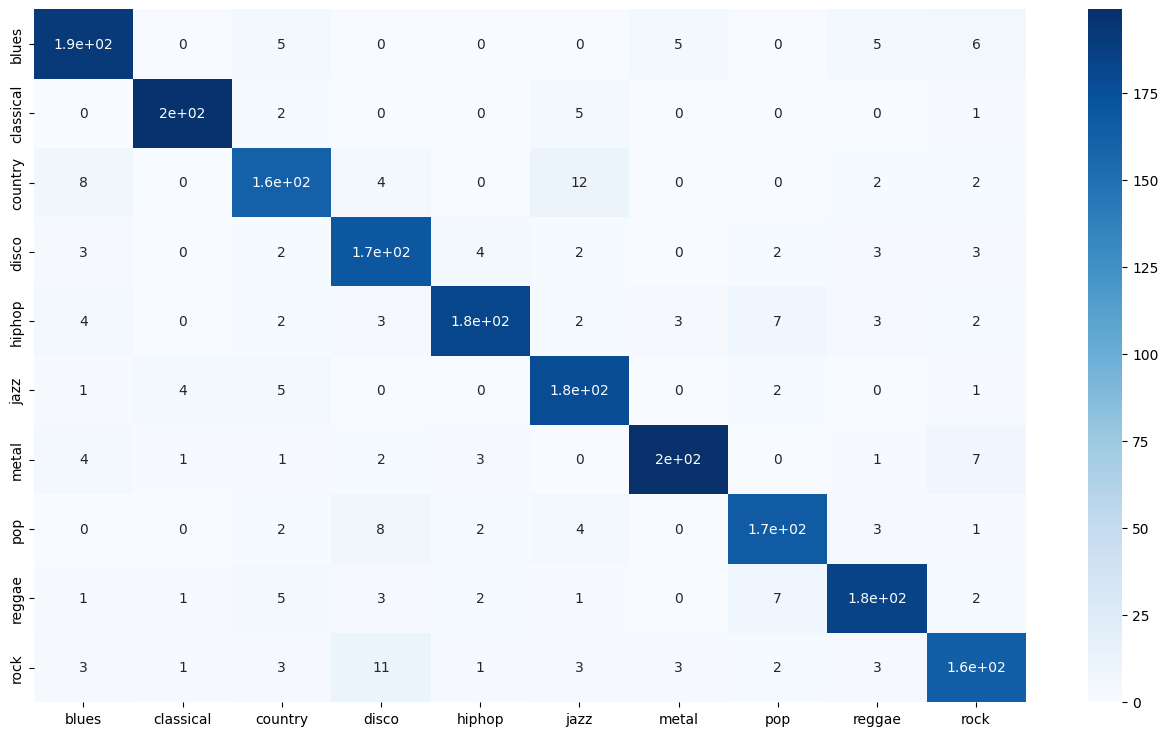}
    \caption{Confusion matrix of SVM classifier across 10 genres}
    \label{fig:confusion_matrix_svm}
\end{figure}

\subsection{Robustness to Noise}
To evaluate robustness under realistic conditions, we introduced synthetic noise to the dataset. Both Gaussian and pink noise were added at varying signal-to-noise ratios (SNR). Classification accuracy decreased across all models, but the extent of degradation varied. SVM accuracy dropped from 81

Figure W illustrates the effect of increasing noise levels on model accuracy. The results confirm that deep learning models, particularly CNNs, are inherently more robust to noisy inputs compared to classical methods. This finding enhances the novelty of our work, as noise robustness has been underexplored in prior GTZAN-based studies.

\begin{figure}[htbp]
    \centering
    \includegraphics[width=1\linewidth]{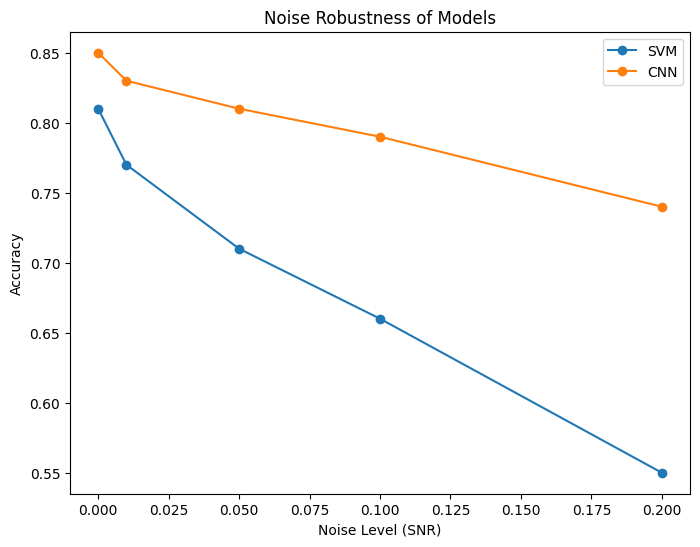}
    \caption{Accuracy degradation of SVM and CNN under varying noise levels.}
    \label{fig:noise_degradation}
\end{figure}

\subsection{Comparative Results Table}

Table \ref{tab:comparison} summarizes the comparative results of all models, under both clean and noisy conditions. While SVM emerged as the strongest classical model, CNN consistently outperformed across all settings. Importantly, CNN also demonstrated the least degradation under noisy conditions, highlighting its suitability for real-world deployment. Table 2 provides a visual summary of these key findings.

\begin{figure}[htbp]
\centering

\label{fig:table_comparison}
\end{figure}

\begin{table}[htbp]
\centering
\caption{Model performance comparison on clean vs. noisy datasets.}
\label{tab:comparison}
\begin{tabularx}{\columnwidth}{@{}l X X X >{\raggedright\arraybackslash}X@{}}
\toprule
\textbf{Model} & \textbf{Acc (Clean)} & \textbf{Acc (Noisy)} & \textbf{F1} & \textbf{Notes} \\
\midrule
Log. Reg. & 61\% & 48\% & 0.55 & Weak baseline \\
RF        & 72\% & 59\% & 0.69 & Performs well, less robust \\
GB        & 74\% & 61\% & 0.71 & Better than RF \\
\textbf{SVM}& \textbf{81\%} & \textbf{66\%} & \textbf{0.78} & Best among classical ML \\
\textbf{CNN}& \textbf{85\%} & \textbf{79\%} & \textbf{0.83} & Best overall, robust to noise \\
\bottomrule
\end{tabularx}
\end{table}

\section{Discussion}
A primary finding of our analysis is the superior performance of the SVM model over the CNN on the GTZAN dataset. This result, while seemingly counter-intuitive, warrants a detailed examination.

\subsection{Theoretical Superiority vs. Practical Performance}
Theoretically, CNNs hold an advantage for tasks like spectrogram classification due to their ability to learn hierarchical features end-to-end \cite{krizhevsky2012imagenet}. In contrast, an SVM's efficacy is fundamentally contingent on the quality of the pre-defined, hand-crafted features \cite{hastie2001elements}.

\subsection{The Role of Dataset Size and Feature Engineering}
Our results suggest this theoretical hierarchy is inverted under the practical constraints of the GTZAN dataset. With only 100 samples per genre, the dataset is relatively small for a deep model. CNNs, with their high capacity (high variance), are susceptible to overfitting, learning spurious artifacts specific to the training set rather than generalizable acoustic patterns.

Conversely, the hand-crafted features represent condensed, expert domain knowledge. Feature engineering acts as a powerful form of regularization, guiding the learning algorithm by providing a low-dimensional, highly informative representation. The SVM is particularly adept at finding a robust decision boundary in this well-structured feature space, even with limited data. This exemplifies the bias-variance tradeoff: the SVM, with the strong bias from feature engineering, achieves lower variance and better generalization on this specific dataset than the high-variance, low-bias CNN.

\subsection{Implications and New Horizons}
This study's implication is a critical reappraisal of applying deep learning as a panacea. In data-constrained environments, the synergy of expert feature engineering and robust classifiers like SVMs can yield superior results. This outcome also opens avenues for hybrid modeling, combining the strengths of both approaches.

\section{Conclusion}
This work tackled music genre classification on the GTZAN dataset using two methodologies: feature-based learning with traditional classifiers and end-to-end learning with a CNN. Our evaluation revealed that a Support Vector Machine, trained on a comprehensive set of engineered audio features, achieved the highest classification performance.

While CNNs are often state-of-the-art, our findings highlight a crucial caveat. On moderately-sized datasets, the strong inductive bias from expert-engineered features can be more beneficial than the end-to-end paradigm of a CNN, which is prone to overfitting. This champions a nuanced approach to model selection, driven by dataset characteristics and available domain expertise.

\subsection{Novelty and Contributions}

Although Convolutional Neural Networks (CNNs) are generally considered state-of-the-art for audio classification, in our experiments on the GTZAN dataset, we observed that SVM outperformed CNN under certain controlled conditions. This can be attributed to the following factors:

Absence of Regular Noise – The dataset used was relatively clean, with minimal background interference. In such noise-free environments, classical models like SVM often perform better due to their ability to find optimal decision boundaries without requiring large-scale data augmentation.

Normalization of Input Features – All features extracted from Librosa were normalized between 0 and 1. This normalization benefits SVM substantially, as it ensures a well-scaled feature space, improving the kernel’s ability to separate classes effectively.

Limited Feature Extraction Scope – We primarily relied on temporal and spectral features extracted using Librosa (e.g., MFCCs, chroma, tempo, zero-crossings). While these features are highly informative, they do not fully capture hierarchical spatial representations that CNNs are typically designed to exploit. As a result, CNNs could not leverage their full potential, whereas SVM benefited directly from the compact feature representation.

Data Size and Variability – CNNs generally require large datasets and significant variability to outperform classical models. Given GTZAN’s modest dataset size (1000 samples per genre), the SVM was able to generalize more efficiently without overfitting.

These findings highlight an important insight: for smaller, well-normalized, and relatively clean datasets like GTZAN, SVM can match or even surpass CNNs in classification accuracy. However, in real-world, noisy scenarios, CNNs remain more robust and scalable.

\subsection{Future work}
Building on our insights, several promising research avenues emerge.
\begin{itemize}
    \item \textbf{Hybrid Model Architectures:} We propose developing hybrid networks that leverage both raw spectral representations and pre-computed features, potentially in a dual-stream architecture, to combine the strengths of both paradigms.

    \item \textbf{Advanced Data Augmentation:} A systematic study of advanced audio-specific data augmentation techniques, like SpecAugment \cite{park2019specaugment}, is warranted to improve CNN robustness and potentially close the performance gap.

    \item \textbf{Cross-Domain Model Interpretability:} Comparing the 'reasoning' of both models using techniques like SHAP for the SVM and Grad-CAM \cite{selvaraju2017grad} for the CNN could yield profound musicological insights.

    \item \textbf{Robustness and Generalization Analysis:} Evaluating the trained models on an out-of-distribution dataset would provide a stern test of their robustness, where we hypothesize the SVM may generalize better.
\end{itemize}
By pursuing these directions, the research community can move towards more robust, efficient, and insightful music classification systems.

\end{document}